\documentclass[final,5p,times,twocolumn]{elsarticle}

\usepackage[utf8]{inputenc}
\usepackage{caption}
\usepackage{amssymb,amsmath}
\usepackage[export]{adjustbox}
\usepackage{float}
\usepackage{tikz}
\usepackage{multirow}
\usepackage{longtable}
\usepackage{listing}
\usepackage{color}
\usepackage{outlines} 
\usepackage{xcolor}

\usepackage{subcaption} 

\journal{Physica A: Statistical Mechanics and its Applications}

\begin{document}

\begin{frontmatter}



\title{Critical phenomena in the market of competing firms induced by state interventionism}


\author{Micha{\l} Chorowski\corref{cor1}%
}
\ead{ma.chorowski@student.uw.edu.pl}
\author{Ryszard Kutner%
}

\cortext[cor1]{Corresponding author}
\address{Faculty of Physics, University of Warsaw, Pasteur Str. 5, PL-02093 Warsaw, Poland}

\begin{abstract}

We achieve two primary goals in this work. First, we propose a flexible algorithm that can simulate various scenarios of state/government intervention. Secondly, we analyze the scenario exhibiting the critical behavior of the market of competing firms, which depends on the degree of government intervention and the activity level of the firms. Thus, we have analyzed the second-order phase transition series, finding the levels of critical intervention and the critical exponent values. As a result, we have observed a sharp increase of fluctuations at a critical intervention level and the local breakdown of the average market technology concerning the frontier technology.  

\end{abstract}



\begin{keyword}
market of competing firms \sep second-order phase-transitions \sep critical phenomena \sep critical threshold of state interventionism \sep critical exponents \sep technology breakdown

\PACS 89.65 Gh \sep 05.40.-a \sep 89.75.Da



\end{keyword}

\end{frontmatter}



\section{Introduction}
With the IT sector, modern production technologies have a decisive impact on the global economy and finance. They have opened a new multidimensional space for competition between firms worldwide. This creates new opportunities for physicists interested in non-linear many-body problems. Recently, Ausloos, Clippe, and P\c{e}kalski have proposed an exciting model \cite{ausloos2004model} later extended by Cichy \cite{cichy}.
It is a non-linear statistical physics simulation model for technological growth, referred to as an ACP-C approach. It is the basis of our work.

Various aspects of growth are still studied in condensed matter, thermodynamics, and statistical physics. It seems natural to extend them to economic growth. The inspiring papers in this respect seem to be refs. \cite{mimkes2012introduction,gatti2005new,mimkes2010stokes,llas2003nonequilibrium,szydlowski2006capital}.

It is commonly believed that mutual penetration technological progress and human capital are of utmost importance for long-term economic evolution. However, there is no consensus on how to take those factors into account in actual models or on the basic mechanisms of technological progress and human capital accumulation \cite{cichy2009human}. One of the most crucial endogenous growth theory models was proposed by Romer \cite{romer1990endogenous,romer1986increasing}, and Lucas \cite{lucas1989mechanics}. Several other models were subsequently proposed \cite{cichy2009human}. 

The model we propose expands upon the ACP-C approach to include the direct  state's influence on market dynamics by using three parameters: the probability of government intervention $0\leq q\leq 1$, intervention efficiency $0\leq \eta \leq 1$, and the probability of the firms' activity $0\leq \lambda \leq 1$. We can look at it as the time-dependent stochastic endogenous and exogenous constraints. 

The government intervention can take the form of stimuli such as tax relief, subsidies, grants, buying a majority stake, or in the extreme case, the nationalization of a firm. Our understanding of state interventionism in the free market is still in its infancy. This problem is as old as the free market itself \cite{aikins2009political,datta1990market,napoles2014macro}.

Many fragmentary studies report the impact of state intervention on GDP (gross domestic product) based on specific examples/situations. 
It was found \cite{barro1990government} that government intervention can have a beneficial influence on endogenous growth.  A study of the postwar period in the United States found that the output increases when government spending increases. However, when taxes increase, output falls \cite{blanchard2002empirical}. It was also found that the relationship between government consumption and GDP growth is negative \cite{grier1989empirical}. A more thorough study claimed that the relationship between government spending and growth could go both ways; it can be both advantageous and disadvantageous \cite{hsieh1994government}. The relationship between government spending and growth can vary across countries, industries, and time.

There is no consensus on the relationship between government spending and economic growth. This is because the relationship can be non-linear, as found by \cite{christie2014effect}. Government spending can be productive or unproductive. The quality of government may also play a vital role \cite{christie2014effect}. The proper model of government intervention should thus allow for both positive and negative effects on output. The type of interventionism that is usually used in practice by governments is another factor relating to specific, real situations. We emphasize that our model only focuses on state/government intervention in the company market and currently does not include other types of interventions such as social transfers, or infrastructure development.

\section{Model}\label{section:model}

Our model is mainly numerical - we define its stochastic dynamics using the non-equilibrium Monte Carlo method, where the detailed balance conditions are not met. We are limited to a variant of the model with a broad spectrum of asymptotic stationary states. This type of model best characterizes the market of competing companies. The local dynamics are defined here using the transition rate, depending on the company's relative technological level. In this way, it introduces positive feedback with the technology of the leader/frontier.

\subsection{Basic definitions}

First, a two-dimensional flat square lattice of size $L \times L$ was created (where $L$ is the linear size of the lattice). Each site/node of the lattice represents a discrete position in which a firm/company can reside. A pair of time-dependent functions $(\omega_i(t),A_i(t))$ describe the state of the node. The first function represents the shares $\omega_i(t)$ of the company on the $i^{th}$ site of the lattice. The second function defines the firm's technology level at this site, $A_i(t)$. An unpopulated lattice site is in the state of $(0,0)$. 

Initially, the lattice is populated with companies according to the specified initial density $c_0$. We achieved it by creating a single company at each lattice site with a probability of $c_0=0.8$ (for reference). Other initial densities are possible. However, the chosen density allows us to compare our results with the corresponding ones of the ACP-C approach \cite{ausloos2004model,cichy,appa}.

The shares $\omega_i(t)$ describe what percentage of the market a given firm at lattice site $i$ covers at any time $t$. The normalization condition
\begin{equation}
\sum_{i=1}^{L^2} \omega_i(t) = 1.
\label{eq:shares}
\end{equation}
holds for any time $t$. The $N(t)$ sites are populated by the firms - the remaining $L^2-N(t)$ sites are empty. 

Initially, we assume a reference egalitarian case that all firms have equal shares $\omega_i(0)=1/ N(0)$, where $N(0)\gg 1$. 

At any time $t$, the system is characterized by the average technology level $\langle A(t) \rangle$, which is the average (expected value) weighted by shares for a single realization of a given Monte Carlo simulation, 
\begin{equation}
\langle A(t)\rangle = \sum_{i=1}^{L^2} \omega_i(t) A_i(t).
\label{eq:Aweight}
\end{equation}

The system starts with an initial average technology $\langle A(t=0) \rangle$, which is the arithmetic average. For the reference purpose we set it to $\langle A(t=0) \rangle = 0.5$. We achieved it by giving all companies a random technology level drawn from a uniform distribution $\mathcal{U}(0,1)$ over the segment $\langle 0,1\rangle $. 

We assume that the progression of technology depends on the global leader's technological growth rate. Companies in different countries can copy the frontier's solutions to increase their technology. It constitutes the technology diffusion stimulated by the outer field. We assume that the leader's technology grows according to Eq. (\ref{eq:frontier}), following Moore's  macroeconomic law of exponential growth \cite{schaller1997moore}. The parameter $\sigma$ is the rate at which the frontier's technology, $F(t)$, increases. 
\begin{equation}
F(t) = F(0)e^{\sigma t},
\label{eq:frontier}
\end{equation}
where we assume for simplicity of further calculations $F(0)=1$. We emphasize that in the context of formula (3), there is extensive literature initiated by Keynes seminal work \cite{JMK}.

\subsection{Algorithm of dynamics}

\subsubsection{Comment on two time scales}

In the Monte Carlo (MC) algorithm, we consider, as usual, two time-scales: (i) Monte Carlo step (MCS) or a single draw (MC second) and (ii) Monte Carlo step per lattice site (MCS/site or MC minute). 

Time scale (i) we use when we consider a single selected company. In this case, the time, $t_{MCS}$, is measured by the total number of draws (seconds). 

In time scale (ii), the situation is more complicated because the number of firms on the lattice, $N(t)$, depends on the time $t$. The point is that each MCS/site should count the same number of MC seconds, regardless of the current number of companies on the lattice. On the other hand, the point is not to randomize empty sites in the lattice, making the algorithm more effective. Therefore, the single MCS/site is equal to the number of lattice sites, with the number of single draws (i.e. the number of MC seconds) equal to the current number of companies, $N(t)$. Thus, to draw randomly $N(t)$ real seconds, we co-opt $L^2 - N(t)$ additional virtual seconds already without drawing. Thanks to this extension, the size of a single MCS/site is constant (regardless of the current number of companies). Thus, in a single MCS/site, each company has (on average) one chance to become active, which is as it should be. Therefore, we treat a single MCS/site as an (egalitarian) time unit, even though it consists of MCSs. That is, it can be divided into smaller time intervals. However, they no longer have the required egalitarian character.
We denote the current number of MCS/site as $t$. So, there is a simple relationship between $t_{MCS}$ and $t$, namely $t_{MCS}=L^2 t+t_{MCS}^{\prime }$, where $0\leq t_{MCS}^{\prime }\leq L^2-1$. As one can see, the time indicated by $t$ is expressed in MCS/site units (MC minutes). On the other hand, the time marked by $ t_{MCS}$, like the time marked by $t_{MCS}^{\prime}$, is expressed in MCS units (MC seconds).

As for the time unit calibration, in \cite{cichy} a single MCS/site corresponded to one year. 
In general, it does not have to be this way - instead of one year, it can be, for example, one month. It depends on the scale (i.e., time horizon) in which the empirical data is collected.

\subsubsection{Different variants of our model}\label{section:variants}

We review our model's basic state interventionism scenarios or variants, which we expect to be (more or less) reflected in reality.

In Variant I (as in all other variants), we use the government intervention probability $0\leq q\leq 1$. When a company is at risk of failure, the government intervenes and saves the company with probability $q$. After being saved, the company does not move on the lattice and cannot perform further actions in the given MCS, $t_{MCS}$. After the intervention, a new MCS begins immediately, and the next company is chosen. The rescued company's temporary passivity can be interpreted as the time required to absorb the help provided. For example, it needs time to reorganize itself internally.

In Variant II, after state intervention with probability $q$, the saved company can move on the lattice and perform subsequent action in the same MCS, $t_{MCS}$. It means that the company was ready to receive support and did not need any time delay to absorb it.

Variants I and II are characterized by interventionism in adverse selection, which occurs when companies will go bankrupt. We considered Variants I and II in \cite{appa}, while here we mention them for completeness.

In Variant III, the state intervention occurs before the survival probability of the company at the $i^{th}$ lattice site, $p_i(t)$, is calculated. It means that the government cannot assess a firm's financial situation accurately. The government's choice of which firms to help is random and independent of the firm's condition - this is an egalitarian approach. 
The government intervenes using the well-known principle of Steinhaus' snapshot observations \cite{HDS}. The opposite case we briefly considered in our earlier work \cite{appa}.

The intervention in Variant III works differently from Variants I and II. In Variant III, we introduce a parameter of intervention efficiency $\eta \in \langle 0, 1\rangle$, which describes the quality of the intervention. In Variant III, the government's intervention with probability $q$ saves the company and can improve its technological level, while the efficiency of this technology improvement is parametrized by $\eta $. This type of technology growth can be interpreted as a grant from the government to fund research and development, purchase advanced technology from other countries, or other similar support activities. After a successful intervention, the next company is picked in Variant III, just like in Variant I.

In Variant IV, interventionism is also egalitarian, as in Variant III. However, after intervention in a given MCS, the firm's activity is random with a probability $0\leq \lambda \leq 1$. Activity in this MCS may end in bankruptcy. 

\begin{figure}
\includegraphics[width=85mm,height=55mm]{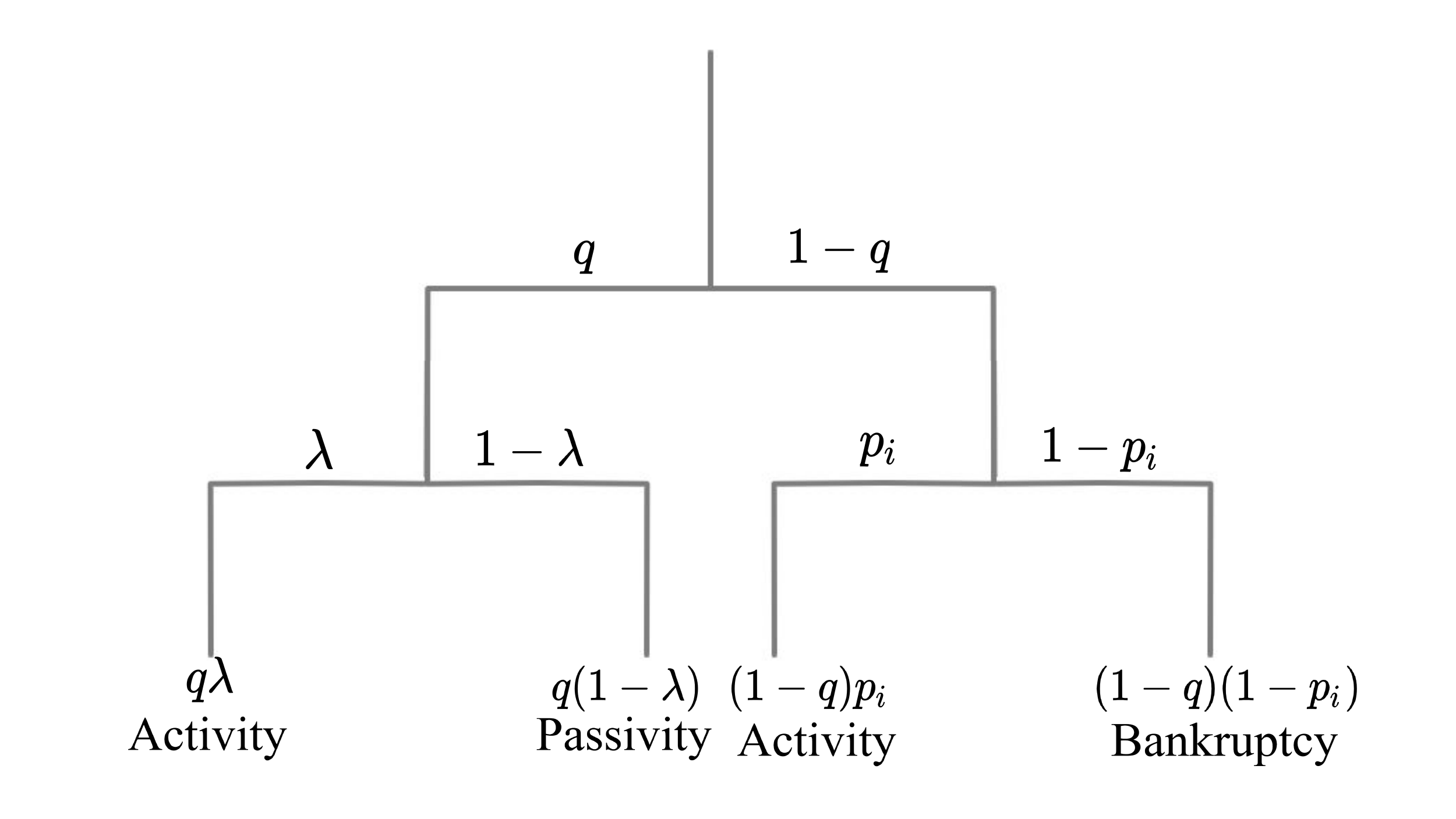}
\caption{The hierarchical binary tree which illustrates the functioning of our reference Variant V for any occupied place $1\leq i\leq N(t)$. Apparently, the local dynamics governed by Eq. (\ref{eq:p}) takes place only after there has been no state intervention. It is described in the right branch of the diagram. The $\eta $ parameter, which is not visible in the diagram, appears in the model at the level of calculating the company's current technological level increase, i.e., immediately after successful intervention when $r_5^{\prime } \leq q$. Thus, it is present implicitly through the survival probability $p_i$.}
\label{fig:DiagramV}
\end{figure}

In Variant V (which is also egalitarian), after successful state intervention with probability $q$ in a single MCS, the company becomes active with probability $\lambda $ (see Fig. \ref{fig:DiagramV}). In this MCS, the company is not threatened with bankruptcy, which is the difference between this more realistic variant and the previous one.
Hence, the combined probability of a company's activity on this single path is $q\lambda $ (on average). In contrast, the company is passive with probability $q(1-\lambda )$.

If the intervention did not occur with probability $(1-q)$, then the company survives at the $i^{th}$ lattice site with probability $p_i$. The combined probability of the company's survival on this alternative path is $(1-q)p_i$ (it is also the probability of the company's activity in a given MCS in the presence of no state intervention). Thus, the total probability of the company's survival in a given MCS is $q\lambda + q(1-\lambda) + (1-q)p_i$ (on average). 
Otherwise, despite intervention, the company goes bankrupt with probability $(1-q)(1-p_i)$ (on average). Note that this probability that a company will go bankrupt plus the above probability that it survives adds up to 1, as it should be due to normalization of probabilities.

The above variant of the model describes situations where, in the absence of state support, the company is forced to compete on the market with other companies. Of course, in such a situation, the company is exposed to bankruptcy. On the other hand, state support allows the company to survive and operate (or not) according to its discretion. This model is egalitarian, i.e. each company (regardless of its status) has an equal chance of obtaining state support. It seems that this type of approach should be favored in states of global crisis, during which time it is about saving the market as a whole without interfering with individual companies' competitive abilities. 

It is essential that we modeled government interventionism and introduced the effectiveness of this interventionism in Variants III - V. This has a direct impact on the current level of technology of any company. 

We expect Variants I - V to be a useful tool package for analyzing companies' real evolution in the market. However, as the reference variant, we choose Variant V, showing that critical phenomena are possible. It is one of the two variants that have this crucial property. The second is Variant II, which seems to be less realistic because it does not allow the state to improve firms' technology, and only weak firms are the target of the intervention. Variant II is a particular case of Variant V.

\subsubsection{How the algorithm works}\label{section:algorwork}

Here, we show how the algorithm works, which enables the system/market evolution within the five variants of our model. 
The graphs in Fig. \ref{figure:4graphs} (in \ref{append:graph-tab}) are helpful in this respect.

The algorithm works as follows:
\begin{outline}[enumerate]
 \1 It randomly picks a single company from $N(t)$ firms. Let the firm be placed at site $i$.
 \1 It calculates the probability of the firm's survival $p_i(t_{MCS})$ (defined by the equations in (\ref{eq:p}) below).
    \2 Variants I and II go directly to item 3 below.
    \2 Variant III takes a random number $r_3\in \langle 0,1\rangle $ drawn from a uniform distribution. If $r_3 \leq q$, the firm's technology increases by quantity $r_3^{\prime } \eta [F(t)-A_i(t_{MCS})]$ (where $r_3^{\prime }$ is a similar random number), ending the current Monte Carlo step, $t_{MCS}$. If $r_3 > q$ then it goes to item 3 below.  
    \2 Before we move on to Variant IV, we will first consider our reference Variant V. A number $r_5^{\prime }$ is drawn. If the intervention is successful, i.e., $r_5^{\prime }\leq q $, then the firm's technology increases by quantity $\tilde{r}_5\eta [F(t)-A_i(t_{MCS})]$ (where $\tilde{r}_5$ is a similar random number). Moreover, the number $ r_5^{\prime \prime }$ is drawn. If $r_5^{\prime \prime }\leq \lambda $, then the algorithm goes to item 4 below. Otherwise, i.e. when $r_5^{\prime \prime }> \lambda $, the company becomes passive in the current MCS and the algorithm returns to the starting point, i.e. item 1, beginning a new MCS. If the intervention is unsuccessful, i.e. $r_5^{\prime } > q $, then a number $r_5^{\prime \prime \prime }$ is drawn. If $r_5^{\prime \prime \prime }\leq p_i$, the algorithm goes directly to item 4 below. Otherwise, i.e. when $r_5^{\prime \prime \prime }> p_i$, the company goes bankrupt (disappears from the lattice) and the algorithm returns to the starting point, i.e. item 1, beginning a new MCS.
    \2 Variant IV differs from the above only in that if $r_5^{\prime \prime }\leq \lambda $ then the algorithm goes to item 3 instead of item 4 below, but an updated (due to intervention with parameter $\eta$) probability of survival, $p_i$, is used. 

 \1 We compare the value of $p_i(t_{MCS})$ with a random number $r_1 \in \langle 0, 1\rangle $ drawn from a uniform distribution. If $r_1 > p_i(t_{MCS})$, the firm goes bankrupt and disappears from the system, leaving its site empty; simultaneously, the current $t_{MCS}$ ends. The shares of the bankrupted firm are equally distributed among all the other firms in the market (egalitarian choice) so that the normalization condition (\ref{eq:shares}) holds. For Variants I and II, when a firm should go bankrupt because $r_1 > p_i(t_{MCS})$, another random number $r_2 \in (0, 1)$ is generated and compared with the parameter $q$. If $r_2 \leq q$, the firm is saved by the government and survives this step despite the fact that $r_1 > p_i(t_{MCS})$. If $r_2 > q$, it means that the intervention was ineffective and the firm goes bankrupt anyway. 
	\2 Variant I: if the government save the company from bankruptcy, the Monte Carlo step $t_{MCS}$ ends. 
	\2 Variant II: if the government save the company, the algorithm goes to item 4 and immediately performs further actions as described below.
\1 The firm attempts to move to one of the four randomly chosen neighboring sites. If the chosen site is occupied, then it does not move and stays in its place. However, it interacts with the firm in that chosen occupied site, as described below in item 6.  
\1  If the randomly chosen site is empty, the firm moves there and checks if this new lattice site has the nearest-neighbor site occupied by any other firm (this neighborhood consists of four lattice sites). If not, then the firm's technology level grows according to the formula $A_i(t_{MCS}+1) = A_i(t_{MCS}) + r_5[F(t)-A_i(t_{MCS})]$, where $r_5\in \langle 0, 1\rangle $ is a random number drawn from a uniform distribution. This growth corresponds to external/exogenous technology diffusion. The company copies the technology of the world frontier/leader imperfectly. The random number $r_5$ is responsible for this imperfection.
\1 There are two mechanisms of inner/endogenous technology diffusion/spreading: (i) the merging of two firms or  (ii) creating a new firm or spin-off. So, as described in item 4, the company located at the $i^{th}$ lattice site moves to a new (empty) site. If the company located at the $i^{th}$ lattice site finds (after moving to a new empty site) a randomly chosen firm at neighboring site $j$, then with probability $b$, the firm at site $i$ merges with the firm at site $j$. The firm at site $j$ disappears from the lattice - it may be, for example, a hostile takeover of a competing company or simply the purchase of a business. The technology of the firm at the $i^{th}$ site changes according to the relation $A_i(t_{MCS}+1) = \textrm{max}[A_i(t_{MCS}), A_j(t_{MCS})]$. The firm at the $i^{th}$ site also takes over the shares of the merged firm, so new $\omega_i$ equals old $\omega_i$ plus old $\omega_j$.

Otherwise (instead of merging), the firms at sites $i$ and $j$ create together with probability $1-b$ a spin-off -- a new firm at $k$ lattice site. In this case, none of the firms disappear from the system/lattice. Site $k$  of the firm is chosen randomly from the nearest and next-nearest neighboring sites of the firm at site $i$. The spin-off appears only if $k$ site was initially empty. The spin-off's technology level is $A_k (t_{MCS}+1) = max[A_i (t_{MCS}), A_j (t_{MCS})]$ and the shares are $\omega_k = \omega_s (\omega_i + \omega_j)$, where $\omega_s \in \langle 0,1\rangle $. Let us add that $\omega_s $ as a model parameter, can also be a random number (and not just fixed as it is now).  Because the normalization condition (\ref{eq:shares}) has to be met, the shares of the company located at $i$ decrease by $\omega_i \omega_s$ and of the company at site $j$ by $\omega_j \omega_s$.
\1 The algorithm returns to item 1 until $N(t)$ firms have been chosen, which ends the current Monte Carlo step/site. 
\end{outline}

Notably, the choice to distribute the company shares after bankruptcy given in item 3 of the algorithm, is not the only possible choice. However, this egalitarian choice seems to be a natural reference case because it does not assume that any companies are privileged in accessing freed market shares. The redistribution of shares is a result of the disappearance of one of the firms from the market, and is not due to competition for the newly available shares. The type of shares redistribution assumed here allows for a comparison of our results with that of work \cite{cichy}. However, the shares could also be distributed differently. For example, the biggest companies could get the most shares according to some preferential selection rule, i.e. after another firm's bankruptcy, according to the principle, that "the rich get richer".

\subsubsection{Probability of firms' survival}

The probability of a firm's survival is an essential element in defining the local dynamics of companies. It takes the form  originally used by the ACP-C approach \cite{ausloos2004model,cichy},
\begin{align} \label{eq:p}
p_i(t_{MCS}) = \left\{ \begin{array}{ll} 
                e^{-sG_i} & \hspace{-1mm}\textrm{if}~G_i>0,~\langle A(t_{MCS}) \rangle < F(t=0) \\
                 1 & \hspace{-1mm}\textrm{if}~G_i\leq 0,~\langle A(t_{MCS})\rangle < F(t=0) \\
                e^{-sH_i} & \hspace{-1mm}\textrm{if}~H_i>0,~\langle A(t_{MCS}) \rangle \geq F(t=0) \\
                1 & \hspace{-1mm}\textrm{if}~H_i\leq 0,~\langle A(t_{MCS})\rangle \geq F(t=0),
                \end{array}\right.
\end{align}
where $G_i=\frac{\langle A(t_{MCS}) \rangle }{F(0)}F(t) - A_i(t_{MCS})$ and $H_i=F(t) - A_i (t_{MCS})$.
Parameter $s=-\partial \ln p_i/\partial G_i$ for the first option and $ s=-\partial \ln p_i/\partial H_i$ for the third option, describes the susceptibility of the market to survival. 

If $s=0$, then the firms never go bankrupt. 
When the $s$ parameter increases, the probability of firms' survival decreases. Hence, a higher $s$ leads to faster technological growth of the system because the system gets rid of technologically weak companies. As a result, $\langle A(t_{MCS})\rangle \geq F(t = 0)$ for a long time and then local system dynamics are mainly governed by  $H_i$. Thus, multi-branch Eq. (4) can be regarded as a preferential selection rule. 

\subsubsection{Free parameters of the model}

The model has free parameters defining the local dynamics: the world technological frontier progress rate $\sigma$, the probability of mergers $b$, technological backwardness susceptibility $s$, the fraction of shares spin-offs inherited from their creators $\omega_s$,  the intervention probability $q$, and its efficiency $\eta$, as well as the probability of a firm's activity $\lambda $. Moreover, the model meets an initial condition, i.e. initial firm concentration $c_0$, and the boundary constraint over time, i.e. the minimal number of companies required for the market to act, $N_{min}$, and a spatial constrain, that is the finite linear lattice size $L$ with a blocking boundary condition. Obviously, $c_0=N(t=0)/L^2$ and  $c_{min}=N_{min}/L^2$. Notably, when $N(t)=N_{min}$, then bankruptcy is forbidden.

For reference purposes we chose $L=40$, $c_0=0.8$, $c_{min}=0.1$, $\sigma=0.01$, $s=1$, $b=0.01$, and $\omega_s=0.1$. All of the above parameter values (except the linear lattice size $L$, which is larger herein) we took from \cite{cichy}. The above parameter values were used in \cite{cichy}, and optimization over $s$ was done to make agreement with the empirical technology data for 27 OECD countries, using the USA as a technological frontier.

\section{Results and discussion}

\subsection{Relaxation of concentration of firms versus stationary states}

\begin{figure}[t]
\includegraphics[width=80mm,height=120mm]{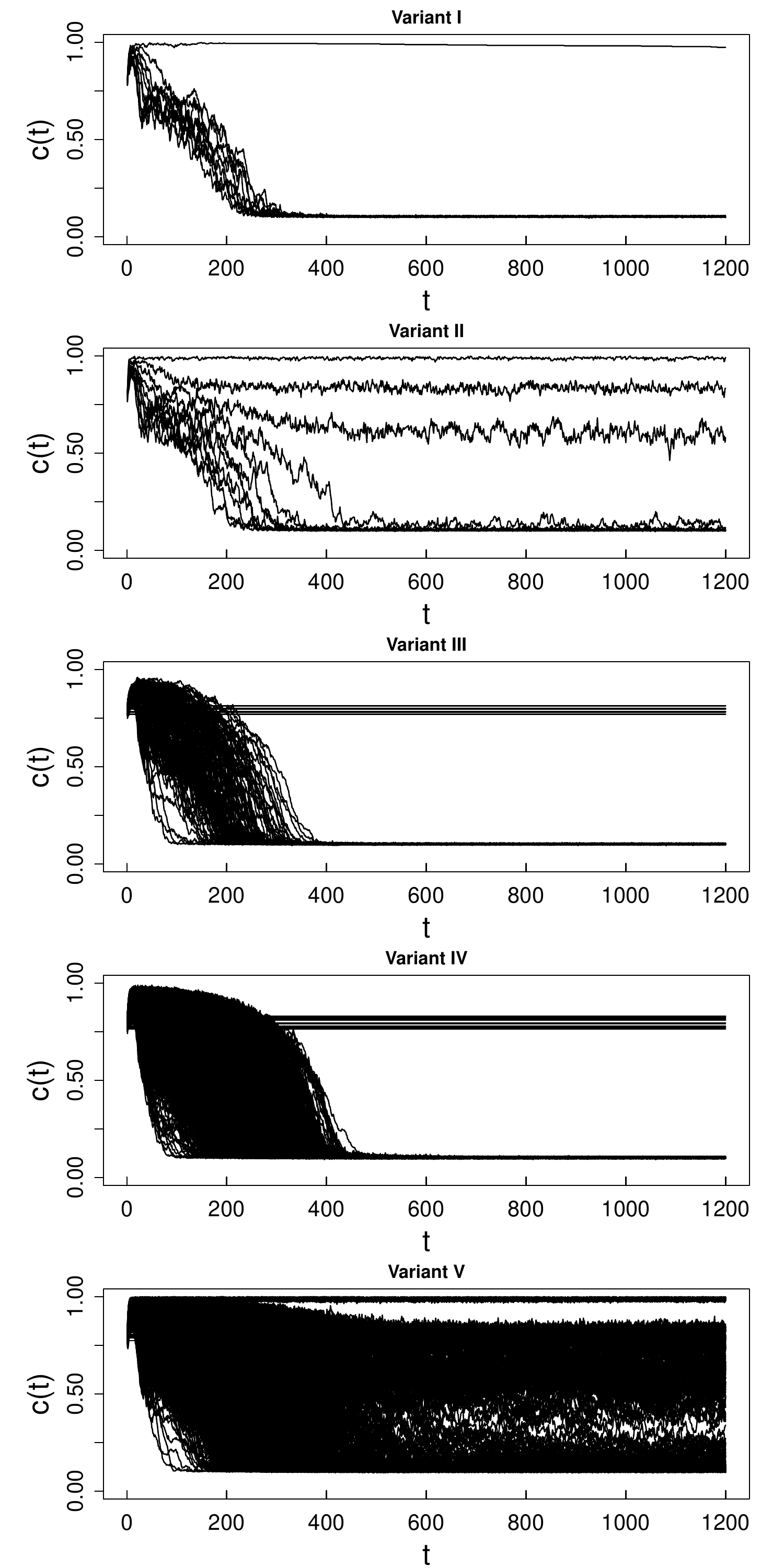}
\caption{Relaxation of the concentration of firms, $c(t)=N(t)/L^2$, for five variants of our model. 
The values of basic parameters are as follows: $L=10,~c_0=0.8,~c_{min}=0.10 ,~\sigma = 0.01,~s=1,~b=0.01,~ \omega_s=0.1$. They are used in subsequent calculations and simulations except size $L$, while the $q,~\lambda $, and $\eta $ parameters were selected from a cube with a side length of $1.0$. As one can see, Variant V has the broadest spectrum of steady states. This is one of the 
reasons for choosing this variant for further study.}
\label{figure:relaxation}
\end{figure}

We present in Fig. \ref {figure:relaxation} the relaxation of the concentration of firms $c(t)=N(t)/L^2$ on the market for our variants of the model. Apparently, the concentration reaches a plateau, i.e. a steady/stationary state, at around 600 MCS/site. We choose Variant V as a reference case because it is the most general containing significant $q, \lambda $, and $\eta $ parameters.

This study examines the properties of stationary states and, in particular, the stationary concentration of firms, $c(t) = c_{st}$, in the market. We treat the concentration of companies $c_{st}$ as a principal order parameter in our model. We consider $c_{st}$ as a function of $q, \lambda $ and $\eta $ at fixed basic parameters $L, c_0, c_{min}, \sigma, s, b$, and $\omega _s$. 

Anticipating the results, we can say that we found critical phenomena in the market of competing companies, which is the main achievement of this work.

\subsection{Criticality in Variant V}\label{section:criticality}

\subsubsection{Critical behavior of $c_{st}$}\label{section:Nst}

\begin{figure}[t]
\includegraphics[width=80mm,height=60mm]{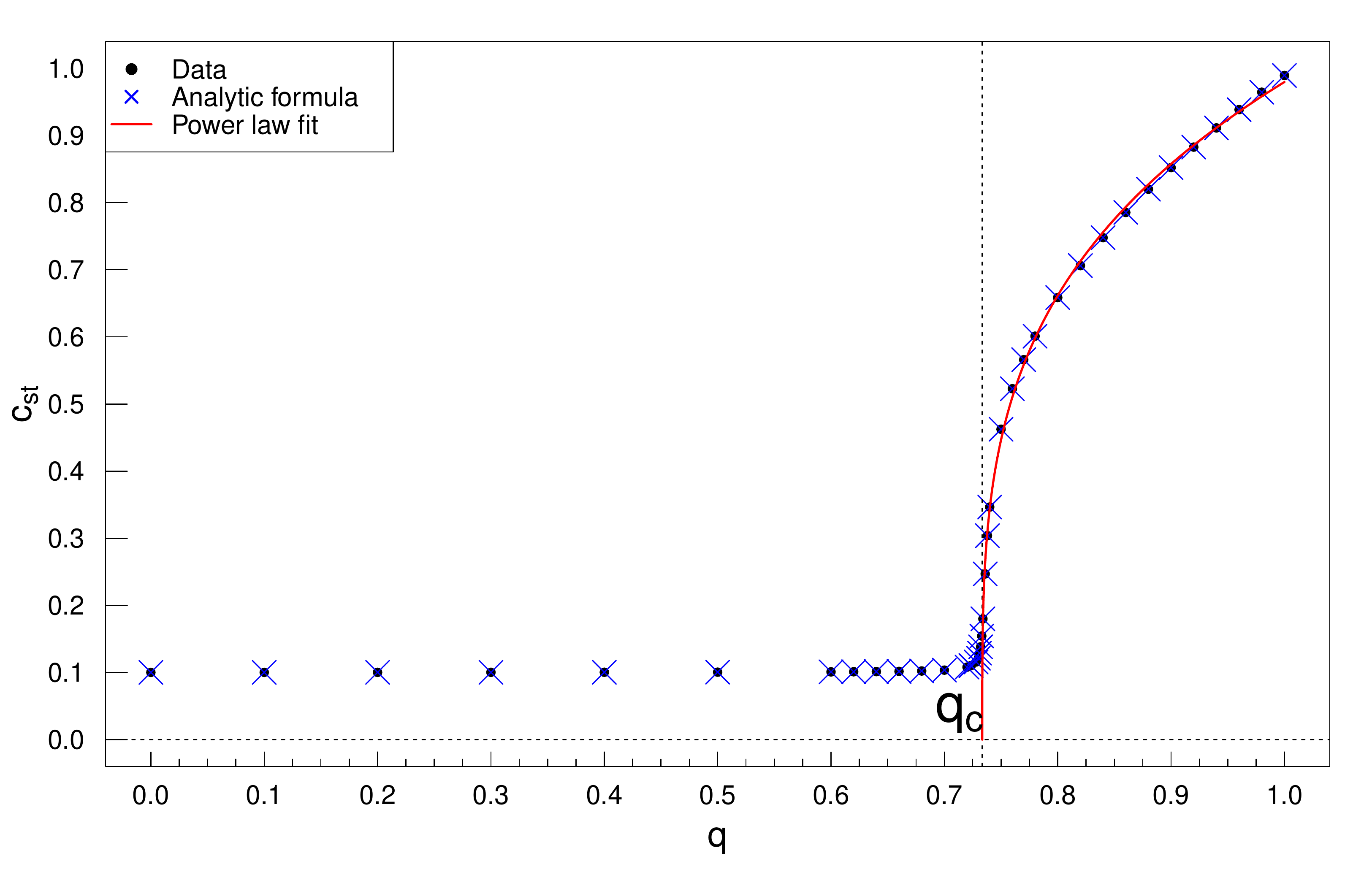}
\caption{Comparison of data for simulation concentration  $c_{st}^{simul}$ vs $q$ (black small circles) with  theoretical $c_{st}^{theor}$ power-law  given by Eq. (\ref{rown:power-law_Nst})  (solid curve), which characterizes a critical behaviour. The following values of the parameters 
of the formula (\ref{rown:power-law_Nst}) were obtained: $A=0.977,~ q_c=0.734$, and $\beta =0.276$, with standard deviations in a third place after the decimal point. The graphic sign x represents the solution given by Eq. (\ref{rown:solutbal}). The dashed vertical line is located at $q_c$. The basic parameters' values here are the same as those used in Fig. \ref{figure:relaxation} except for $L = 40$.}
\label{figure:Nstql09e05}
\end{figure}

Fig. \ref{figure:Nstql09e05} shows the dependence of the stationary concentration of companies, $c_{st}=N_{st}/L^2$, in the market of competing firms obtained by simulation (small circles) versus the probability of state intervention $q$, chosen for example probability values $\lambda = 0.9 $ and $\eta = 0.5$. The solid line represents the prediction of the formula (\ref{rown:power-law_Nst}), while the graphic sign x gives the solution (\ref{rown:solutbal}). 
One can see very good agreement of these three approaches for $q >q_c$. The latter approach is consistent with simulated data over the entire $q$-range. Similar results are obtained for the remaining values of $\lambda > 0$.

The power law, 
\begin{eqnarray}
c_{st}^{theor}=A\cdot \left(\frac{q-q_c}{1-q_c}\right)^{\beta },
\label{rown:power-law_Nst}
\end{eqnarray}
describing the critical behavior with a visible critical value $q = q_c = 0.734$, the critical exponent $\beta=0.257$, and the prefactor $A=0.969$ were adjusted to the simulation results (they statistical errors are at the third place after the decimal point). 

\subsubsection{Abrupt increase of fluctuations}\label{section:fluct}

\begin{figure}
\includegraphics[width=80mm,height=60mm]{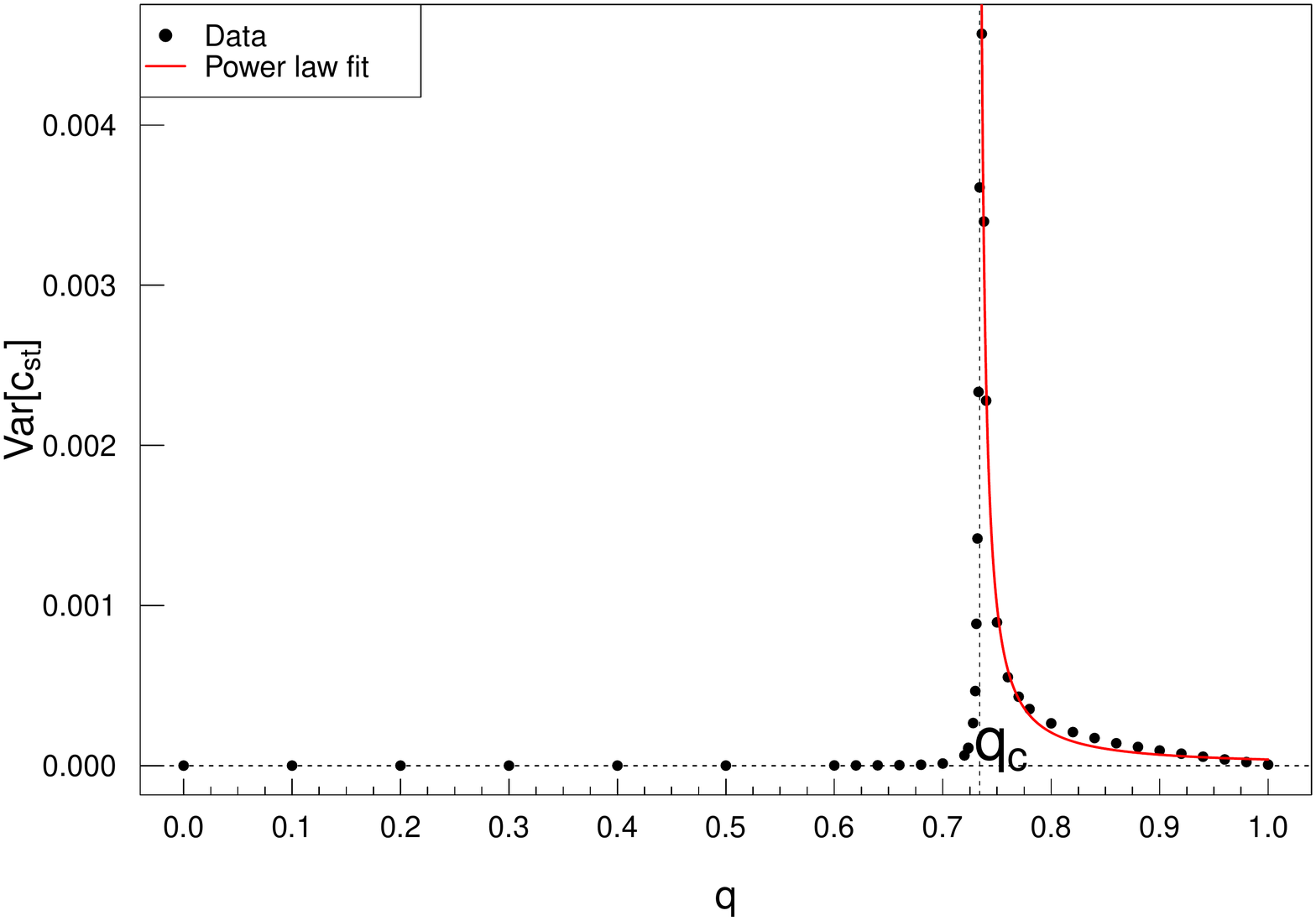}
\caption{Comparison of the empirical variance of $c_{st}^{simul}$ vs $q$ obtained from simulation or numerical experiment (circles) with a theoretical variance (solid curve) given by a power-law describing the critical behavior (see Eq. (\ref{rown:power-law_Var}) for details). From the fit, we obtained the following parameter values: $B=3.8~10^{-5},~q_c=0.730$, and $\gamma =1.247$. The basic parameters' values here are the same as those used in Fig. \ref{figure:relaxation} except for $L = 40$.}
\label{figure:Varql09e05}
\end{figure}

At the critical value $q = q_c$ the fluctuation of $c_{st}$ sharply increases, as we show clearly in Fig. \ref{figure:Varql09e05}. The simulation data (or data from numerical experiments) for the variance presented there are described for $q > q_c$ very well by the power law given by a formula analogous to Eq. (\ref{rown:power-law_Nst}),
\begin{eqnarray}
Var(c_{st}^{theor})&=&\langle \left(c_{st}^{theor}\right)^2\rangle - \langle c_{st}^{theor}\rangle ^2 \nonumber \\
&=&B\cdot \left(\frac{1-q_c}{q-q_c}\right)^{\gamma },
\label{rown:power-law_Var}
\end{eqnarray}
where critical exponent $\gamma = 1.247~(> 2\beta )$ and prefactor $B=3.83~10^{-5}$ (with standard deviations of these quantities in the third place after the decimal point). So at the critical point, we are dealing with unlimited fluctuations, as might be expected. 

Behind this sharp increase in variance is a sharply growing gap between the number of spin-offs created by firms and the number of firms' mergers and bankruptcies.

\subsubsection{Diverging susceptibility}\label{section:chi}

\begin{figure}
\includegraphics[width=80mm,height=60mm]{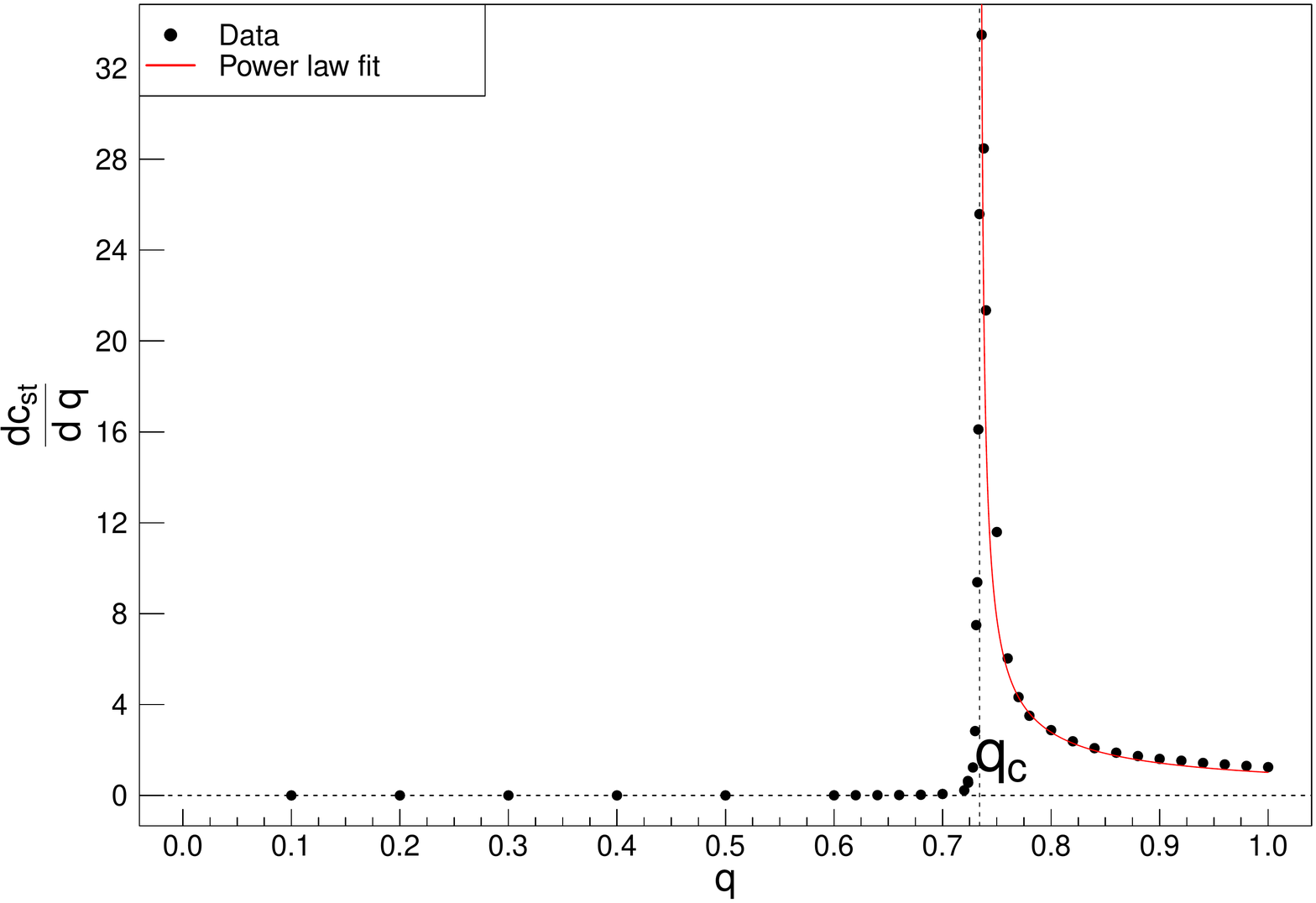}
\caption{Comparison of the susceptibility of $\chi(c_{st}^{simul})$ vs $q$ obtained in numerical experiment (dots) with a theoretical one $\chi(c_{st}^{theor})$ (solid curve) given by a power-law describing critical behavior (see Eq. (\ref{rown:power-law_chist})). As can be seen, this susceptibility creates a divergent peak. Here we used the values of the $A,~q_c$, and $\beta $ parameters that we obtained from the appropriate fitting shown in Fig. \ref{figure:Nstql09e05}. The basic parameters used are the same as for Fig. \ref{figure:Nstql09e05} except for $L=40$.}
\label{figure:chiql09e05}
\end{figure}

The system susceptibility, $\chi $, defined below, diverges for $q\rightarrow q_c$ (see Fig. \ref{figure:chiql09e05} for details). 
\begin{eqnarray}
\chi (c_{st}^{theor})&\stackrel{\rm def.}{=}&\frac{dc_{st}^{theor}}{dq}= \nonumber \\
&=&A\cdot \frac{\beta }{1-q_c}\cdot \left(\frac{1-q_c}{q-q_c}\right)^{1-\beta }.
\label{rown:power-law_chist}
\end{eqnarray}
Formula (\ref{rown:power-law_chist}), resulting directly from formula (\ref{rown:power-law_Nst}), reproduces quite well for $q > q_c$ the data from the simulation, which was to be expected, especially because the parameters $A, q_c$, and $\beta $ were taken from the fit shown in Fig. \ref{figure:Nstql09e05}. The result is, in fact, a test of the self-consistency of our approach. 

All results presented in Secs. \ref{section:Nst}, \ref{section:fluct},  and  \ref{section:chi} directly indicate that we are dealing here with a continuous type (II-order) phase transition and hence with critical phenomena in the market of competing companies.

\subsection{From the balance condition to critical behaviour}\label{section:balaance}

The stationarity of the number of companies on the market of competing companies results from the balance condition that the system can achieve. This condition says that as many companies appear on the market per unit of time as go bankrupt. This is a general balance condition, not a detailed balance condition. Therefore, this condition  leads (in general) to a statistical steady state and not to a statistical equilibrium. 

\begin{figure}[t]
\centering
\advance\leftskip-0cm
\includegraphics[width=80mm,height=60mm]{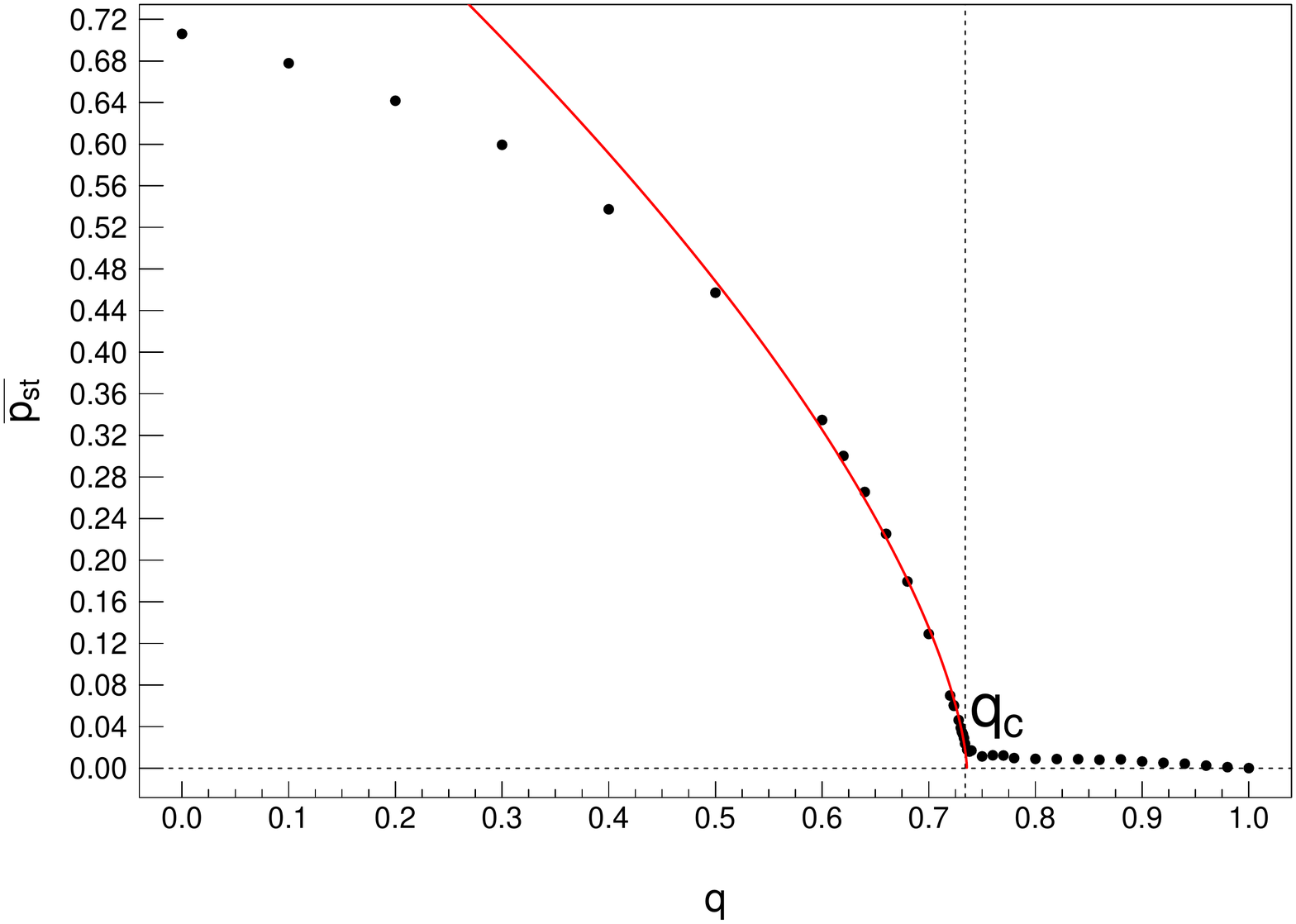}
\caption{Probability $\bar{p}_{st}$ vs $q$. The characteristic shape of simulation (dots) and theoretical (solid) curves is the result of the phase transition at the critical threshold $q=q_c$. The vertical dotted line is located at this threshold. The solid curve we calculated using formula (\ref{rown:arithpst}), while the dots were obtained from simulation.}
\label{fig:pst}
\end{figure}

Using the diagram shown in Fig. \ref{fig:DiagramV}, we have presented in Table \ref{table:stationar} the gain and loss probability currents, which are related to Variant V.
\begin{table*}[t]
    \centering
    \caption{Gain and loss probability currents}
    \begin{tabular}{c|c}
    \hline 
      Gain probability current& Loss probability current \\
      \hline \hline
      $G(1)=z_1(z-1)q\lambda (1-b)f c_{st}(1-c_{st})$  &  $L(1)=z_1q\lambda bc_{st}$ \\
      \hline
      $G(2)=z_1(z-1)(1-q)\bar{p}_{st}(1-b)f c_{st}(1-c_{st})$ & $L(2)=z_1(1-q)\bar{p}_{st}bc_{st}$ \\
      \hline
       ---  &  $L(3)=(1-q)(1-\bar{p}_{st})$ \\
        \hline
    \end{tabular}
    \label{table:stationar}
\end{table*}
We have adopted the following notations here: $z_1$ is the number of the nearest neighbors and $z$ is the total number of the nearest- and next-nearest neighbors. In the case of a square lattice (used in our simulations), $z_1=4$ and $z=z_1+z_2=8$, where $z_2$ is the number of next-nearest-neighbors. The expressions in the table depend at most on the size of the first and second coordination zones and not on the type of lattice/network. In this sense, they are universal. Table \ref{table:stationar} allows the balance condition in the following form,
\begin{eqnarray}
&&G(1)+G(2)=L(1)+L(2)+L(3) \Longleftrightarrow \nonumber \\
&\Longleftrightarrow &(z-1)f(1-b)c_{st}^2 - [(z-1)f(1-b)-b]c_{st}+ \nonumber \\
&+&\frac{1}{z_1}\frac{(1-q)(1-\bar{p}_{st})}{q\lambda+(1-q)\bar{p}_{st}}=0,
\label{rown:balance}
\end{eqnarray}
which gives a quadratic equation for variable $c_{st}$. The probability $\bar{p}_{st}$ is defined by the following arithmetic average,
\begin{eqnarray}
\bar{p}_{st}=\frac{1}{N_{st}}\sum_{i=1}^{N_{st}}p_i(t),
\label{rown:arithpst}
\end{eqnarray}
where the index "$st$" refers to the stationary situation, i.e. for long enough time $t$ such that $p(t), c(t)$, and $\langle A (t)\rangle /F(t)$ reach a plateau (i.e. their stationary values $\bar{p}_{st}, c_{st}$, and $\{\langle A (t)\rangle /F(t)\}_{st}$, respectively). Moreover, the summation in the formula above means the sum of only those lattice sites that are occupied by companies, and when $N(t)=N_{min}$ we take $p_i(t)=1$ because bankruptcy is not permitted in that case. The dependence of $\bar{p}_{st}$ on variable $q$ at fixed $\lambda =0.9$ and $\eta =0.5$ is shown in Fig. \ref{fig:pst}. The solid curve we derived from the formula analogous to that given by the Eq. (\ref{rown:power-law_Nst}). Namely, for $q<q_c$ we have $\bar{p}_{st}=A\cdot (q_c-q)^{\alpha }$, where $A=1.213$, $\alpha = 0.658$, and $q_c=0.736$ were obtained from the fit to data within the scaling region. Thus, we obtained three critical exponents, of which $\alpha $ governs the phase with a lower level of state intervention, while $\beta $ and $\gamma $ governs the phase with a higher level of state intervention. Both phases are related to the common critical threshold of  $q=q_c$. 

\begin{figure}[t]
\centering
\advance\leftskip-0cm
\includegraphics[width=80mm,height=60mm]{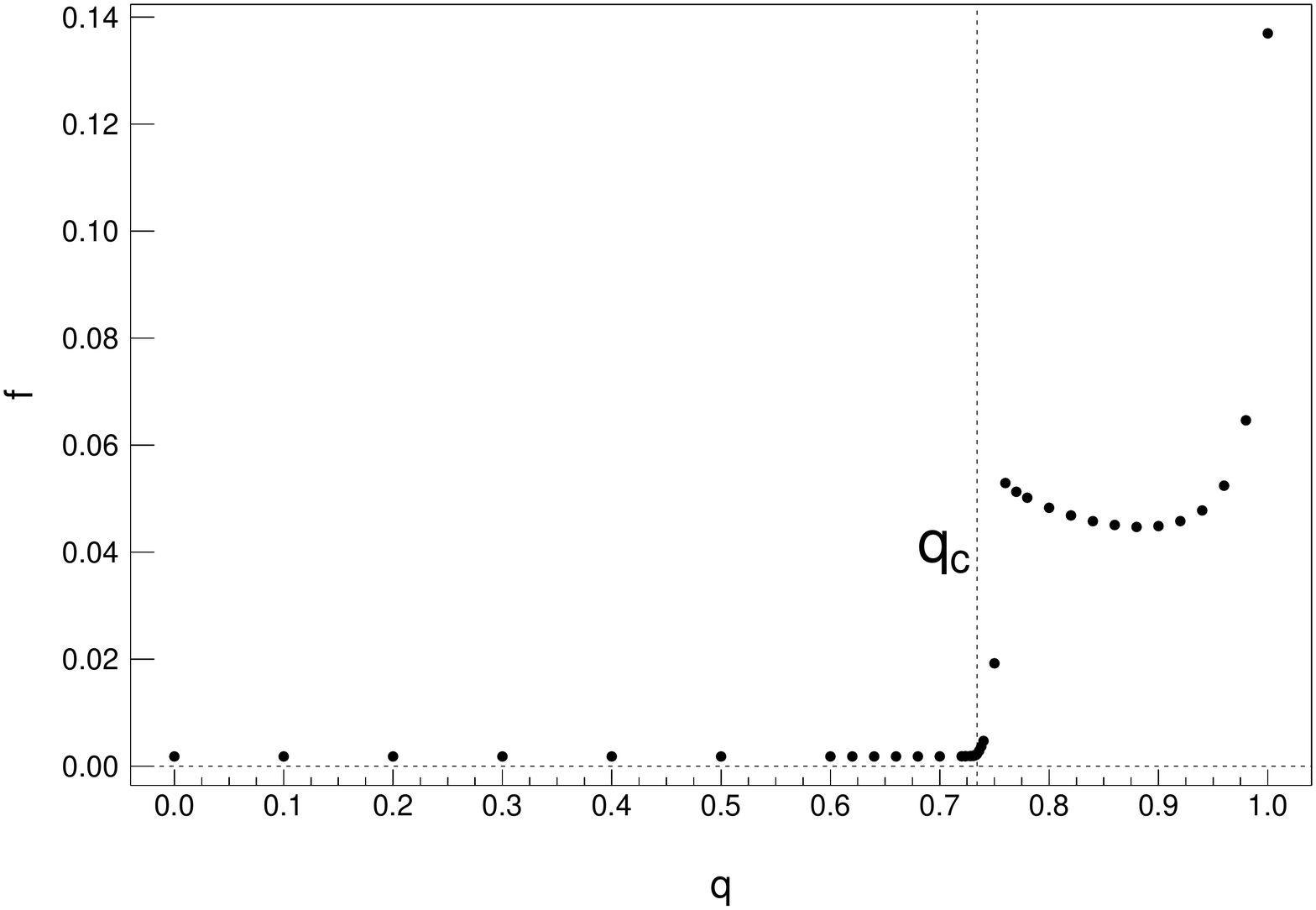}
\caption{Statistical independence factor $f$ vs $q$. The characteristic steep increase in $f$ is clearly visible at $q=q_c$. The case of $q=1$ is different because it concerns the situation where three (partial) probability currents (containing the factor $1-q$) disappear (see Table \ref{table:stationar} for details).}
\label{fig:flambda}
\end{figure}

The statistical independence factor $f$ (defined by Eq. (\ref{rown:f}), which indicates the lack of coupling between companies located in the neighboring lattice sites and the vacancy (empty lattice site), depends on $q, \lambda $, and $\eta $. Its non-monotonic dependence on $q$ at fixed $\lambda =0.9$ and $\eta =0.5$ is shown in Fig. \ref{fig:flambda}. Factor $f$ was calculated numerically by using self-consistently solution (\ref{rown:solutbal}).
Two different behaviors depending on whether $q$ is less than or greater than $q_c$, are observed.

Equation (\ref{rown:balance}) has a solution of the form,
\begin{eqnarray}
c_{st}^{\mp }&=&\frac{(z-1)f(1-b)-b}{2(z-1)f(1-b)}\cdot (1 \nonumber \\
&\mp &\sqrt{1-\frac{(z-1)f(1-b)}{[(z-1)f(1-b)-b]^2}\frac{(1-q)(1-\bar{p}_{st})}{q\lambda+(1-q)\bar{p}_{st}}}). \nonumber \\
\label{rown:solutbal}
\end{eqnarray}
We chose the solution $c_{st}^+$ because the second solution, $c_{st}^-$, vanishes at $q=1$ and is, therefore, 
not compatible with simulations. The solution $c_{st}^+$
is shown in Fig. \ref{figure:Nstql09e05}: the fit to the simulation data is very good over the full $q$ range.

The special case of $q=1$ is particularly simple because it concerns the situation where three components (containing the factor $1-q$) disappear (see Table \ref{table:stationar} for details). The two other non-vanishing components lead directly to the simplified formula, $c_{st}=1-\frac{1}{(z-1)f}\frac{b}{1-b}$, which we also get from Eq. (\ref{rown:solutbal}).


\subsection{Technology evolution}

Now, we focus on the mechanisms of technology growth in the model. The possible mechanisms of market technology change in Variant V are as follows: (i) bankruptcy of technologically weak companies, (ii) mergers of companies, (iii) creating spin-offs, (iv) copying the technology of the leader, and (v) copying due to the government's support with effectiveness $\eta $.
The mechanism that most significantly influences the technology level in the long run is copying the leader's technology (\ref{eq:frontier}).

The technology evolution (in Variant V) of each company due to government's direct help occurs in item 2 subitem (c) of the algorithm. through the $\eta $ parameter.

\begin{figure}[t]
\centering
\advance\leftskip-0cm
\includegraphics[width=80mm,height=60mm]{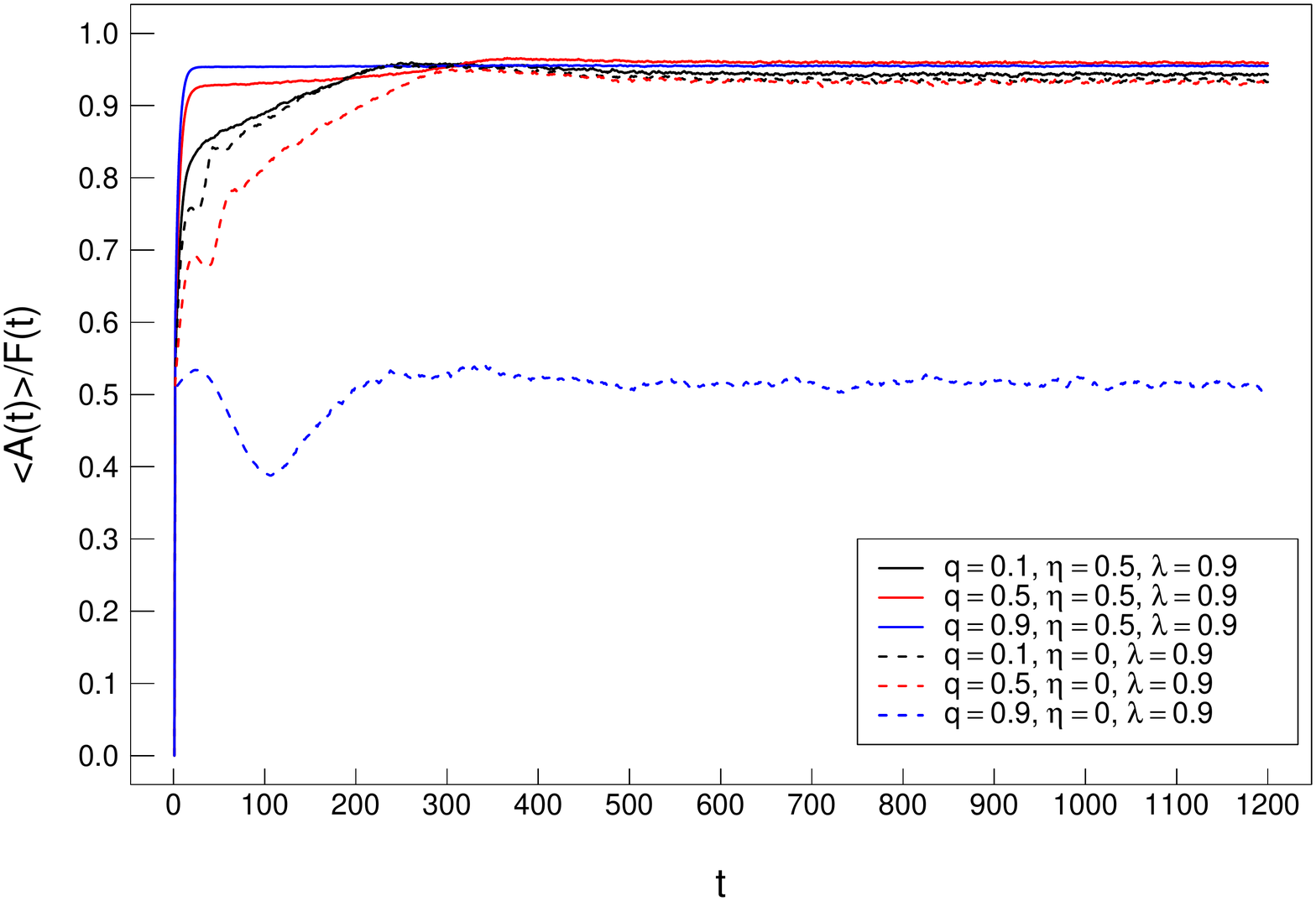}
\caption{The ratio $\langle A(t)\rangle /F(t)$ vs time obtained from simulations for different sets of parameters $q, \eta $ and $\lambda $. The results are averaged over 400 independent replicas or simulations. (The values of basic parameters are the same as for Fig. \ref{figure:relaxation}.)
}
\label{fig:alt4_trajecexample}
\end{figure}

\begin{figure}[tp]
\includegraphics[width=80mm,height=55mm]{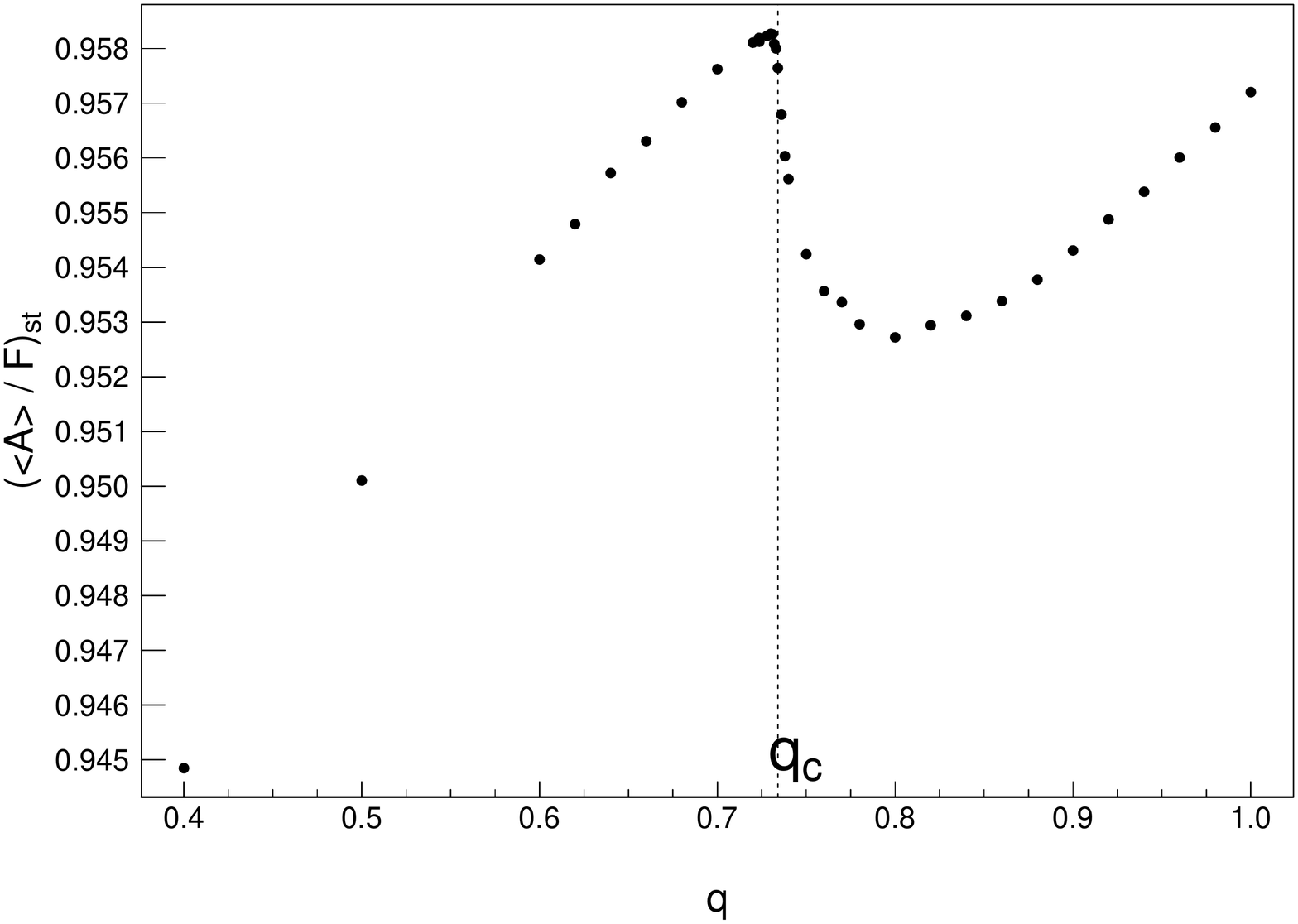}
\caption{Dependency of simulation $\left(\langle A (t)\rangle /F(t)\right)_{stat}$ vs  $q$. Its breakdown is clearly visible at $q=q_c$. Hence, susceptibility $d\left(\langle A (t)\rangle /F(t)\right)_{stat}/dq$ drops at this point. One can see that this drop has a finite amplitude. (The values of basic parameters are the same here as for Fig. \ref{figure:relaxation}.)}
\label{figure:AtFt}
\end{figure}

We examine the steady/stationary state of the relative average level of technology $\langle A (t)\rangle /F(t)$. Fig. \ref{fig:alt4_trajecexample} shows the time dependence of this quantity. 
It reaches a steady-state already for $t$ at about $600$ MCS/site. This is about the relative average level of technology reaching a plateau and not the average technology itself. 
The plateau height of $\langle A (t) \rangle /F(t)$ is controlled not only by the $q$ parameter but also by the $\eta $ parameter. 
Furthermore, in Fig. \ref{figure:AtFt}, we show that the relative technology dips suddenly at $q_c$. Approaching the critical point from the left leads to higher relative market technology, but crossing it might lead to an abrupt decrease forming a local minimum. Hence, the analyzed market would have the highest technology close to $q_c$. Unfortunately, this would be at the cost of large fluctuations in the number of firms.

\section{Concluding remarks}

We extend the stream of works belonging to the rapidly developing area of diffusion/spreading innovation. We extend this stream with the subject of spreading innovation technologies in the market of competing companies. This extension consists of taking into account state interventionism. We examine the critical phenomena occurring in the steady state accompanying this interventionism.

We chose an egalitarian Variant V of our model for the study, i.e. one in which no company is preferred. As it transpired, this variant has the broadest spectrum of stationary states and leads to critical phenomena depending on the level of intervention, $q$, and company activity level $\lambda $. Variant II also leads to criticality, however, this criticality is a particular case of that of Variant V. It does not allow technology to increase due to the intervention (parameter $\eta $ is not present in Variant II). It is, therefore, less realistic and less flexible. Let us add that Variants I - II use the rule of preferential supporting state intervention in companies threatened with bankruptcy. Our model is also open to other preferential selection rules (e.g. the one in which the most  innovative companies are supported). 

The main result of our work is presented in Fig. \ref{figure:Nstql09e05}. We compare three types of results: simulations, the solution of the balance equation, and the power-law function. This comparison, along with the rest of the results for the unlimited increase in fluctuation and susceptibility at criticality threshold $q=q_c$, clearly indicates the presence of critical phenomena in competing firms' markets. Moreover, we checked that similar results are obtained, for example, for $0< \lambda , \eta \leq 1 $. Thus, both the criticality threshold $q_c$ and the critical exponents $\alpha, \beta $ and $\gamma $ are direct functions of $\lambda $ and indirect functions of efficiency of intervention $\eta $. Our results warn that a transition crossing a critical threshold, from the lower-level phase of state intervention ruled by the critical exponent $\alpha $ to the higher-level phase of state intervention ruled by the exponents $\beta $ and $\gamma $, leads to the emergence of a sharp increase of fluctuations and susceptibility on the market and the accompanying local collapse at $q_c$ of average technology. 

As far as the application of the model presented in the paper is concerned, the analysis of the global subprime crisis in 2007-09, during which the help of states in saving individual banks and the banking sector as a whole was obvious, may raise particular hopes.

Moreover, the potential significance of the model during and after a pandemic stems from the fact that the model describes the impact of state intervention on economic growth, and can therefore be a valuable guide in finding optimal conditions for stimulating the economy. However, this aspect of the model has not been sufficiently documented in the present work and rather expresses our supposition. 

Practical use of the model requires introducing operational definitions of the model parameters and their non-trivial calibration, which we are currently working on. 

We suppose our model can bring a better understanding of the relationship between market dynamics and government intervention. It is a generic non-equilibrium agent-based model, having asymptotic stationary states. It uses the dynamic of individual heterogeneous companies on a micro level to describe the macro level's aggregate output. We mainly deal with asymptotic stationary states and the phase transitions accompanying them, especially with critical behaviors. It allows us to escape the critique oriented on equilibrium macroeconomic models \cite{mccauley2006response, mccauley2004dynamics}. 

\appendix
\section{Graphical summary}\label{append:graph-tab}

\begin{figure*}[t]
\centering
  \begin{subfigure}[b]{0.36\linewidth}
    \centering
    \hspace{-1cm}
    \includegraphics[width=\linewidth,height=32mm]{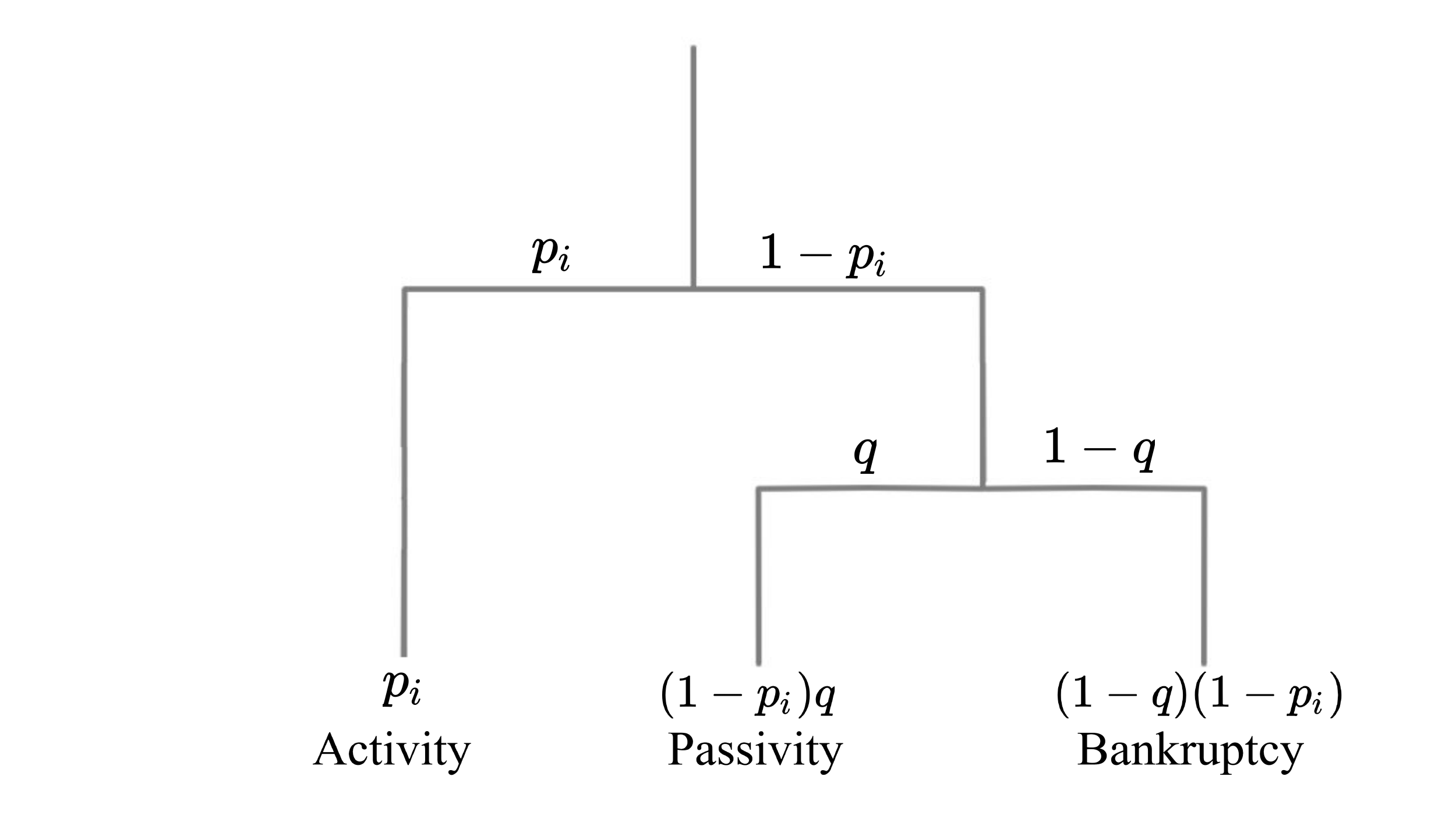} 
    \caption{Variant I} 
    \label{fig7:a} 
    \vspace{4ex}
  \end{subfigure}
  \begin{subfigure}[b]{0.36\linewidth}
    \centering
    \includegraphics[width=\linewidth,height=32mm]{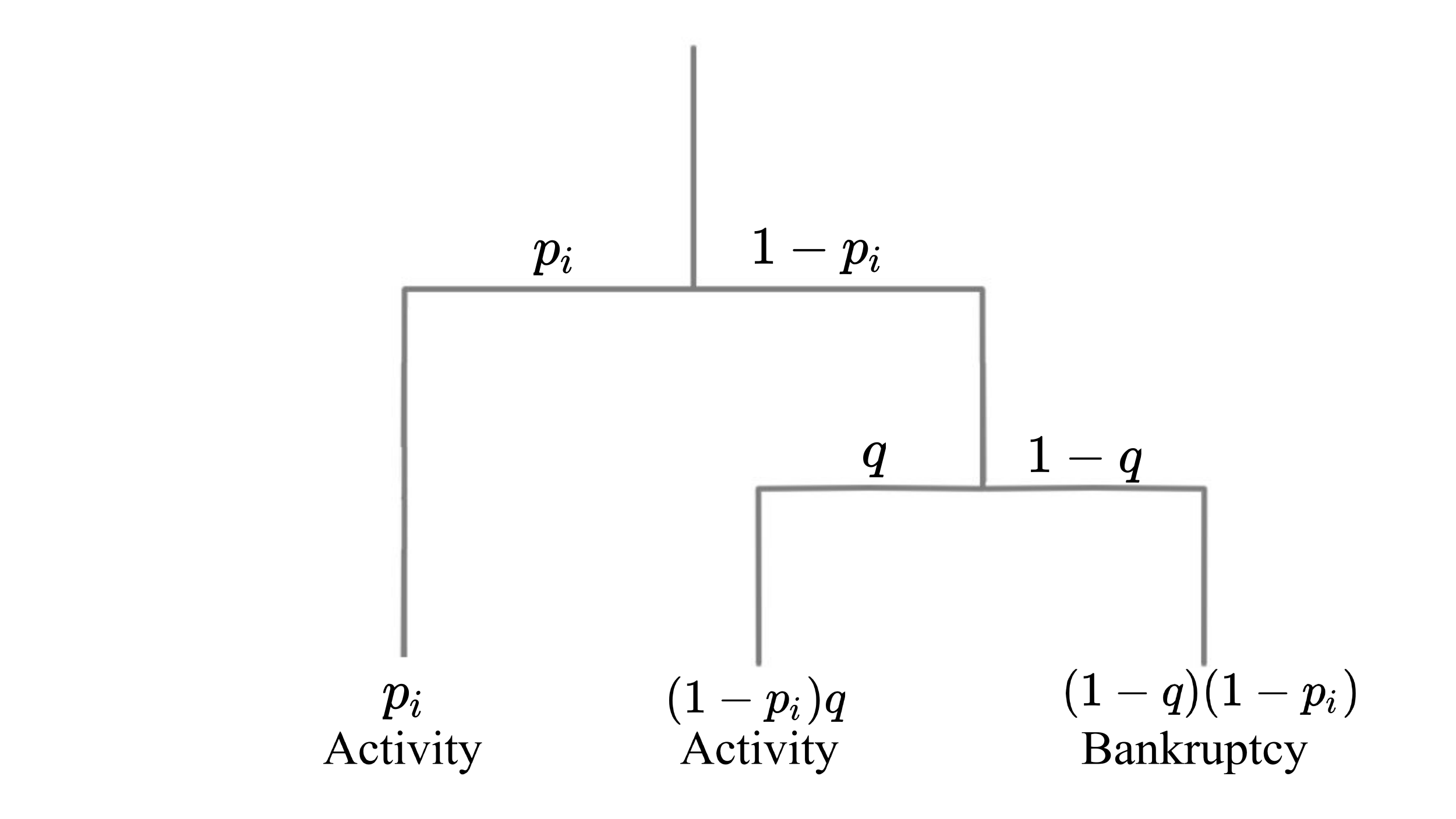} 
    \caption{Variant II} 
    \label{fig7:b} 
    \vspace{4ex}
  \end{subfigure} 
  \begin{subfigure}[b]{0.36\linewidth}
    \centering
    \hspace{-1cm}
    \includegraphics[width=\linewidth,height=32mm]{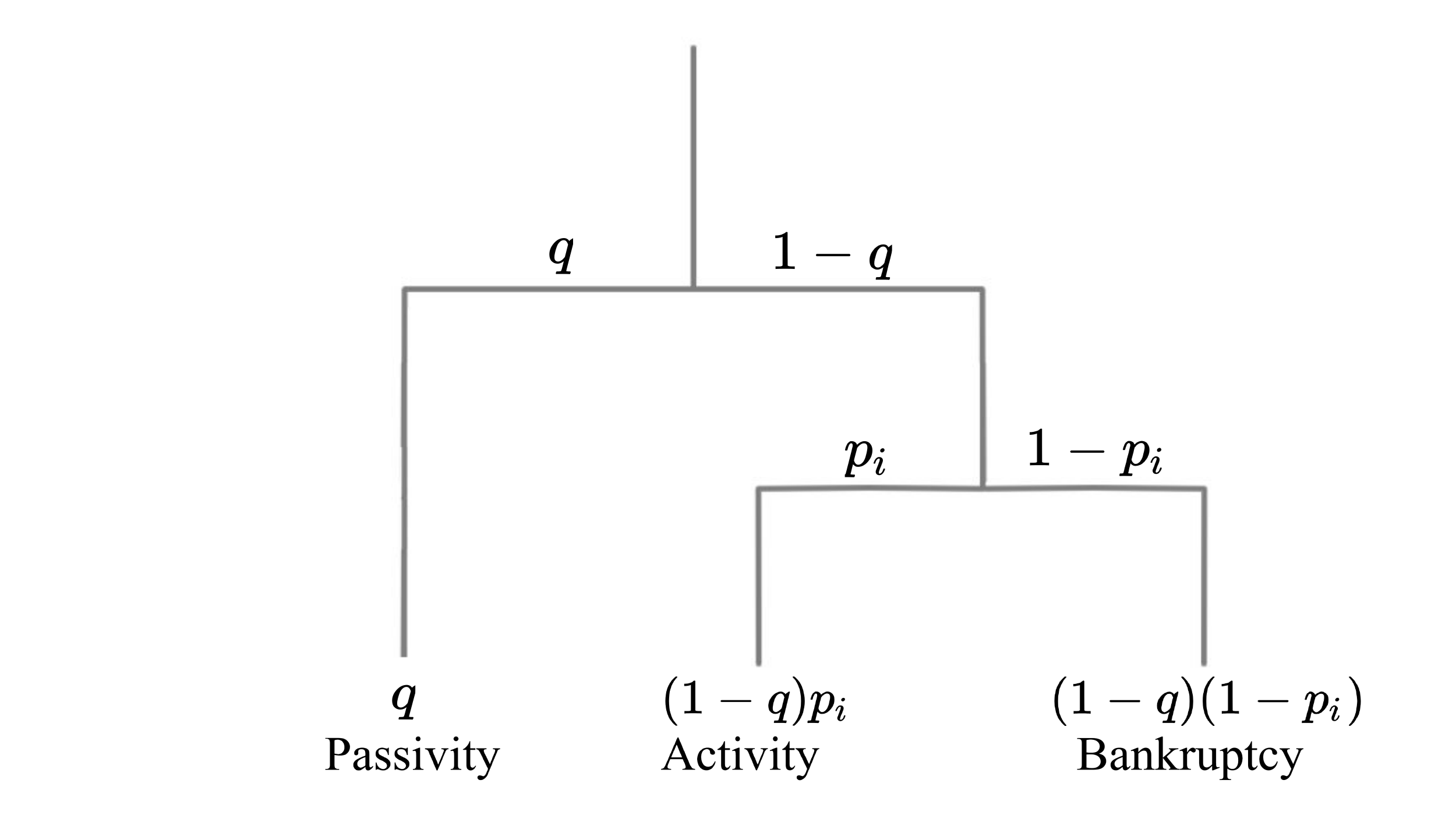}
    \caption{Variant III} 
    \label{fig7:c} 
  \end{subfigure}
  \begin{subfigure}[b]{0.36\linewidth}
    \centering
    \includegraphics[width=\linewidth,height=32mm]{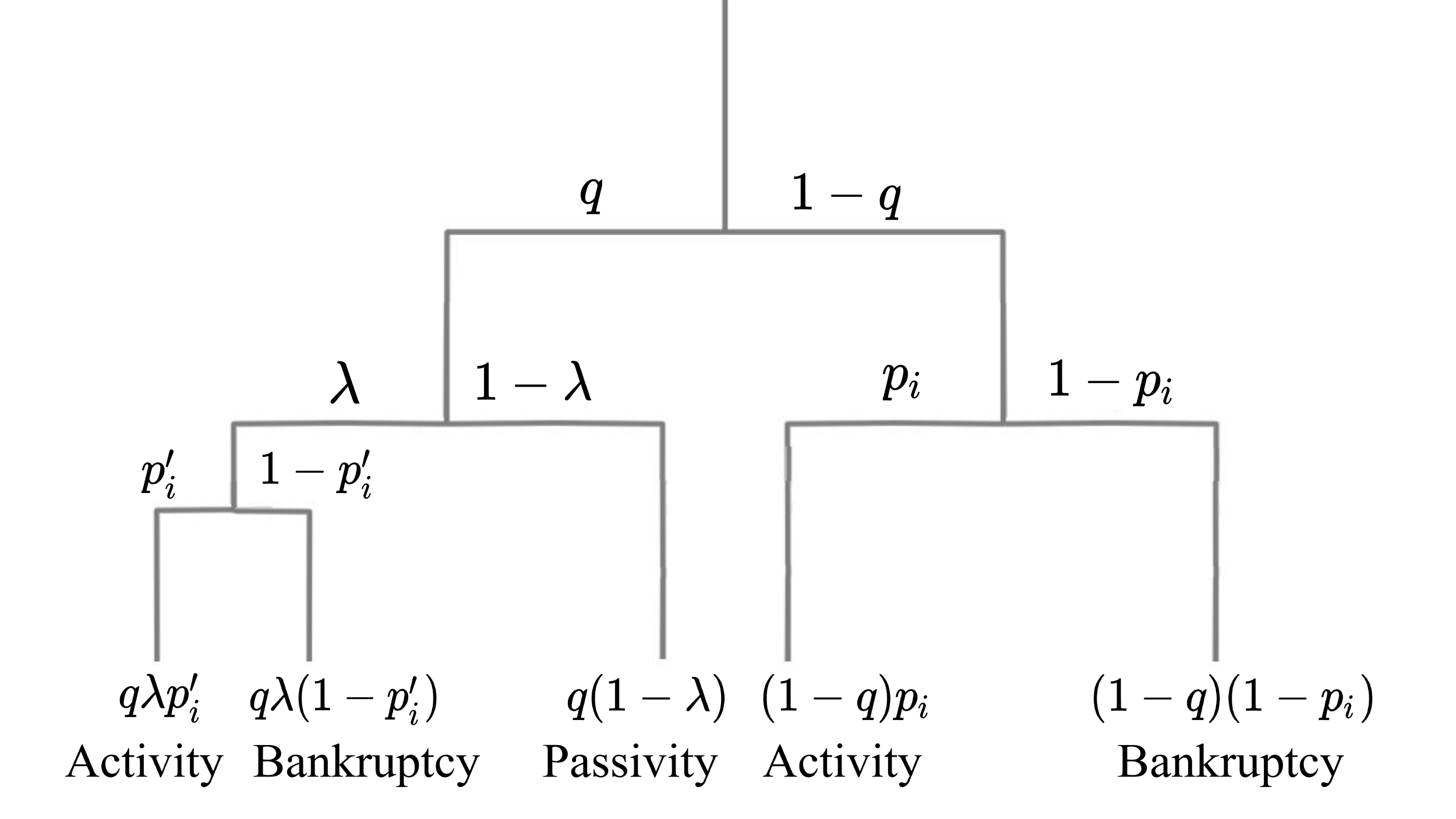} 
    \caption{Variant IV} 
    \label{fig7:d} 
  \end{subfigure} 
  \caption{The hierarchical binary trees of Variants I-IV of our model (analogous to the binary tree shown in Fig. \ref{fig:DiagramV}). For a description of the probabilities used here, see Sec. \ref{section:model}.}
  \label{figure:4graphs} 
\end{figure*}

We present a graphical summary of subsequent variants of our model presented in subsections \ref{section:variants} and \ref{section:algorwork}. As shown in Fig. \ref{figure:4graphs}, only for Variant IV 
the likelihood of activity, passivity, and bankruptcy depend on $\lambda $. The dependence on $\lambda $ for the other three variants appears only at the gain and loss current probabilities, that is, at the level of balance equations for different variants (not shown here). As for Variant V, some likelihood of activity and passivity also depends on $\lambda $ (cf. Fig.\ref{fig:DiagramV}), and the same applies to gain and loss current probabilities (cf. Table \ref{table:stationar}). 
  .

\section{Overview Table I}\label{append:TableI}

\begin{figure}[h!]
\centering
\includegraphics[width=60mm,height=60mm]{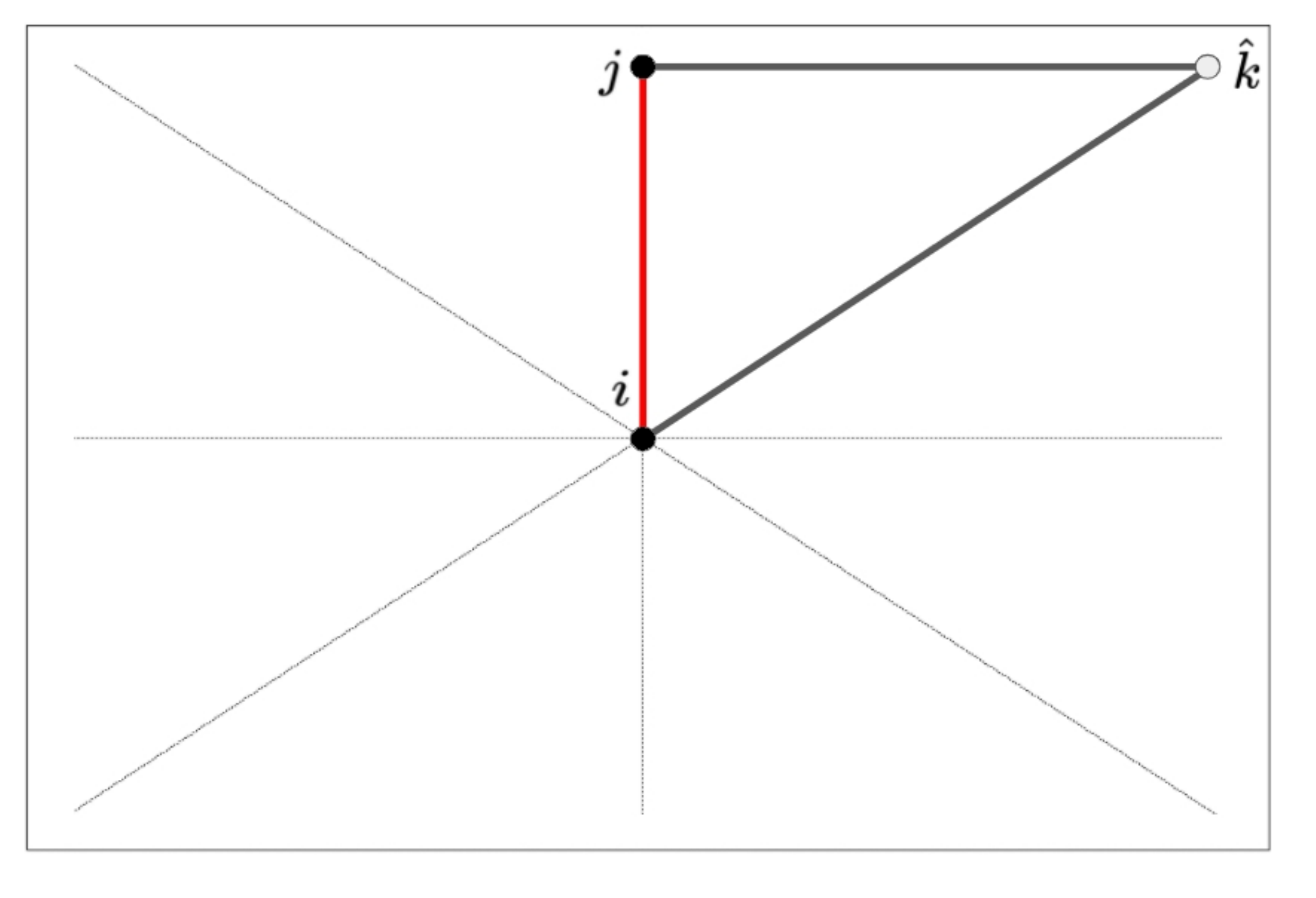}
\caption{The nearest- and next-nearest neighbors of the lattice site $i$. The nearest-neighbor site $j$ is occupied by a company (black circle). The number of $z_1$'s nearest neighbors equals 4.  The pair of $(i, j)$ sites are distinguished by a bold (red) vertical segment, which emphasizes that they remain in the relationship. The number of next-nearest neighbors, $z_2$, is also 4. The vacancy is located in one of these sites $\hat{k}$ (empty circle). It belongs to the next-nearest neighbors of site $i$. This site is related to the pair of sites mentioned above, marked with thin solid lines. The remaining lattice sites surrounding the site $i$ are joined with $i$ by very thin lines.}
\label{figure:Fig_Append}
\end{figure}

We discuss Table \ref{table:stationar} on the basis of Figs. \ref{fig:DiagramV} and \ref{figure:Fig_Append}. 

(i) Element $G(1)$ concerns the most extreme left path of the three. This expression is the partial probability per unit time (MCS/site) or the partial probability of a current of a spin-off creation. We define this current when state intervention occurs, and the company is already active after that in the same MCS.  

The conditional probability $p(j,\hat{k}|i)$ of finding the pair of lattice sites $(j,\hat{k})$ as neighbors of a company located at site $i$ (as depicted in Fig. \ref{figure:Fig_Append}) can be factorized as follows, 
\begin{eqnarray}
p(j,\hat{k}|i)=p(j)p(\hat{k})f,
\label{rown:pjk}
\end{eqnarray}
where a factor of statistical independence,
\begin{eqnarray}
f=\frac{p(i|\hat{k})}{p(i)}\frac{p(\hat{k},i|j)}{p(\hat{k},i)},
\label{rown:f}
\end{eqnarray}
was obtained based on the famous Bayesian theorem. Note that $f=1$ only if the occupied lattice site $i$ and empty lattice site $\hat{k}$ are statistically independent, and similarly the pair of sites $(\hat{k},i)$ are statistically independent of the occupied lattice site $j$. 
The occupation probabilities of $p(i)$ and $p(j)$ can be approximated by $c_{st}$ in the stationary states, and analogously the probability of a vacancy appearing at site $k$ (or $p(\hat{k})$) can be approximated by $1-c_{st}$. 

(ii) The element $G(2)$ 
is related to the third path of the diagram counting from the left side. \ref{fig:DiagramV}. The conditional probability $p(j,\hat{k}|i)$ is given by Eq. (\ref{rown:pjk}) supported by Eq. (\ref{rown:f}).  This element also describes the probability current to create a spin-off but in the absence of state intervention.

(iii) The element 
$L(1)$ defines the probability current of the company's disappearance in a given lattice site as a result of merging it with the company in the neighboring lattice site. This probability current  
relates to the leftmost branch of the tree.

(iv) The element 
$L(2)$ refers to the probability current 
associated with the third-from-the-left branch of the tree. 

(v) The element $L(3)$ refers to the probability current associated with the third-from-the-left branch of the tree. It describes the bankruptcy of a firm at a given site. 



  \bibliographystyle{elsarticle-num} 
\bibliography{bibliografia_new_corr}

\begin{thebibliography}{10}
\expandafter\ifx\csname url\endcsname\relax
  \def\url#1{\texttt{#1}}\fi
\expandafter\ifx\csname urlprefix\endcsname\relax\def\urlprefix{URL }\fi
\expandafter\ifx\csname href\endcsname\relax
  \def\href#1#2{#2} \def\path#1{#1}\fi

\bibitem{ausloos2004model}
M.~Ausloos, P.~Clippe, A.~Pekalski, Model of macroeconomic evolution in stable
  regionally dependent economic fields, Physica A: Statistical Mechanics and
  its Applications 337~(1-2) (2004) 269--287,
  https://doi.org/10.1016/j.physa.2004.01.029.

\bibitem{cichy}
K.~Cichy, Microeconomic evolution model with technology diffusion, Acta Physica
  Polonica A 121~(2B), doi: 10.12693/APhysPolA.121.B-16 (2012).

\bibitem{mimkes2012introduction}
J.~Mimkes, Introduction to macro-econophysics and finance, Continuum Mechanics
  and Thermodynamics 24~(4-6) (2012) 731--737.

\bibitem{gatti2005new}
D.~D. Gatti, C.~Di~Guilmi, E.~Gaffeo, G.~Giulioni, M.~Gallegati, A.~Palestrini,
  A new approach to business fluctuations: heterogeneous interacting agents,
  scaling laws and financial fragility, Journal of Economic behavior \&
  organization 56~(4) (2005) 489--512.

\bibitem{mimkes2010stokes}
J.~Mimkes, Stokes integral of economic growth: Calculus and the solow model,
  Physica A: Statistical Mechanics and its Applications 389~(8) (2010)
  1665--1676.

\bibitem{llas2003nonequilibrium}
M.~Llas, P.~M. Gleiser, J.~M. L{\'o}pez, A.~D{\'\i}az-Guilera, Nonequilibrium
  phase transition in a model for the propagation of innovations among economic
  agents, Physical Review E 68~(6) (2003) 066101.

\bibitem{szydlowski2006capital}
M.~Szydlowski, A.~Krawiec, On capital dependent dynamics of knowledge, arXiv
  preprint physics/0608197 (2006).

\bibitem{cichy2009human}
K.~Cichy, Human capital and technological progress as the determinants of
  economic growth, National Bank of Poland Working Paper~(60), available at
  SSRN: https://ssrn.com/abstract=1752094 or
  http://dx.doi.org/10.2139/ssrn.1752094 (2009).

\bibitem{romer1990endogenous}
P.~M. Romer, Endogenous technological change, Journal of political Economy
  98~(5, Part 2) (1990) S71--S102.

\bibitem{romer1986increasing}
P.~M. Romer, Increasing returns and long-run growth, Journal of political
  economy 94~(5) (1986) 1002--1037.

\bibitem{lucas1989mechanics}
R.~E. Lucas, On the mechanics of economic development, NBER Working
  Paper~(R1176) (1989).

\bibitem{aikins2009political}
S.~K. Aikins, Political economy of government intervention in the free market
  system, Administrative Theory \& Praxis 31~(3) (2009) 403--408,
  https://doi.org/10.2753/ATP1084-1806310309.

\bibitem{datta1990market}
M.~Datta-Chaudhuri, Market failure and government failure, Journal of Economic
  Perspectives 4~(3) (1990) 25--39, dOI: 10.1257/jep.4.3.25.

\bibitem{napoles2014macro}
P.~R. N{\'a}poles, et~al., Macro policies for climate change: Free market or
  state intervention?, World Social and Economic Review 2014~(3, 2014) (2014)
  90.

\bibitem{barro1990government}
R.~J. Barro, Government spending in a simple model of endogeneous growth,
  Journal of political economy 98~(5, Part 2) (1990) S103--S125.

\bibitem{blanchard2002empirical}
O.~Blanchard, R.~Perotti, An empirical characterization of the dynamic effects
  of changes in government spending and taxes on output, the Quarterly Journal
  of economics 117~(4) (2002) 1329--1368.

\bibitem{grier1989empirical}
K.~B. Grier, G.~Tullock, An empirical analysis of cross-national economic
  growth, 1951--1980, Journal of monetary economics 24~(2) (1989) 259--276.

\bibitem{hsieh1994government}
E.~Hsieh, K.~S. Lai, Government spending and economic growth: the g-7
  experience, Applied Economics 26~(5) (1994) 535--542.

\bibitem{christie2014effect}
T.~Christie, The effect of government spending on economic growth: Testing the
  non-linear hypothesis, Bulletin of Economic Research 66~(2) (2014) 183--204.

\bibitem{appa}
M.~Chorowski, R.~Kutner, Government intervention modeling in microeconomic
  company market evolution, Acta Physica Polonica A (in print)Proceedings of
  the 10th Polish Symposium of Physics in Economy and Social Sciences FENS, 3-5
  July 2019, Otwock - Swierk. ArXiv preprint arXiv:2007.06451 (2020).

\bibitem{schaller1997moore}
R.~R. Schaller, Moore's law: past, present and future, IEEE spectrum 34~(6)
  (1997) 52--59, doi: 10.1109/6.591665.

\bibitem{JMK}
J.~M. Keynes, The general theory of employment, The quarterly journal of
  economics 51~(2) (1937) 209--223.

\bibitem{HDS}
H.~Steinhaus, The problem of estimation, The Annals of Mathematical Statistics
  28~(3) (1957) 633--648.

\bibitem{mccauley2006response}
J.~L. McCauley, Response to “worrying trends in econophysics”, Physica A:
  Statistical Mechanics and its Applications 371~(2) (2006) 601--609.

\bibitem{mccauley2004dynamics}
J.~L. McCauley, Dynamics of markets: econophysics and finance, Cambridge
  University Press, 2004.

\end{thebibliography}

\end{document}